\documentclass[a4paper]{article}

\usepackage{INTERSPEECH_v2}
\usepackage{hyperref}
\usepackage[british]{babel}  
\usepackage[T1]{fontenc} 
\usepackage{fix-cm}
\title{Attention and Localization based on a Deep Convolutional Recurrent Model for Weakly Supervised Audio Tagging}
\name{Yong Xu, Qiuqiang Kong, Qiang Huang, Wenwu Wang, Mark D. Plumbley}
\address{
Center for Vision, Speech and Sigal Processing, University of Surrey, UK}
\email{\{yong.xu, q.kong, q.huang, w.wang, m.plumbley\}@surrey.ac.uk}

\begin{document}

\maketitle
\begin{abstract}
Audio tagging aims to perform multi-label classification on audio chunks and it is a newly proposed task in the Detection and Classification
of Acoustic Scenes and Events 2016 (DCASE 2016) challenge. This task encourages research efforts to \text{better} analyze and understand the content of the huge amounts of \text{audio} data on the web. The difficulty in \text{audio} tagging is that it only has a chunk-level label without a frame-level label. This paper presents a weakly supervised method to not only predict the tags but also indicate the temporal locations of the occurred acoustic events. The attention scheme is found to be effective in identifying the important frames while ignoring the unrelated frames. The proposed framework is a deep convolutional recurrent model with two auxiliary modules: an attention module and a localization module. The proposed algorithm was evaluated on the Task 4 of DCASE 2016 challenge. State-of-the-art performance was achieved on the evaluation set with equal error rate (EER) reduced from 0.13 to 0.11, compared with the convolutional recurrent baseline system.
\end{abstract}
\noindent\textbf{Index Terms}: audio tagging, attention model, DCASE 2016 challenge, convolutional recurrent model

\section{Introduction}

Environmental audio processing is gaining increasing research interest following the large amount of work on speech and music processing. The 1st DCASE challenge (DCASE 2013) focused on audio scene and event recognition \cite{giannoulis2013detection, stowell2015detection}.
The 2nd DCASE (DCASE 2016) challenge \cite{dcase2016} introduced a new task, namely audio tagging \cite{dcase_t4, foster2015}. Audio tagging mainly aims at determining the presence of events in the acoustic scene. Meanwhile, localizing the acoustic events that have occurred would also be interesting but difficult considering that the label is in chunk-level rather than frame-level. The chunk-level labeled data was regarded as weakly labeled data \cite{kumar2016audio}.

The traditional method for audio tagging is based on Gaussian mixture model (GMM) trained on Mel frequency cepstrum coefficients (MFCCs) \cite{Foster2016, Yun2016}. Since the introduction of the DCASE 2016 challenge, many deep learning based methods have been developed for audio tagging. Deep neural network (DNN) has been used to predict the audio tags \cite{xu2016fully, Kong2016}. Different from the GMM method, the DNN-based method can model all of the tags in shared weights simultaneously. However, convolutional neural network (CNN) was shown to perform better than the DNN \cite{Cakir2016, Lidy2016}.
Currently, the best performing system was introduced in \cite{yong2017ijcnn} where \text{convolutional} gated recurrent neural network incorporating spatial features was adopted.
However all of these methods can not locate the occurred acoustic events in the audio chunk. Acoustic event localization based on weakly labeled data will be one focus of this paper. Multiple Instance Learning based event detection \cite{kumar2016audio} is a related method which was adopted for weakly labeled data. On the other hand, most of the training of above neural network models were actually ill-posed due to a context window input (e.g., 32 frames or 640 miliseconds in \cite{Cakir2016}) which only represents the partial segment of the whole chunk. Nonetheless, the given label is in chunk-level. This assumes that the chunk-level label is also matched on the partial segment which is not always reasonable.

Recently, attention-based neural networks have been applied to a wide variety of tasks, such as speech recognition \cite{bahdanau2016end, chorowski2015attention}, visual object classification \cite{mnih2014recurrent}, machine translation \cite{bahdanau2014neural} and image caption \cite{xu2015show}. We use the term {\em \textbf{attention}} to mean to focus on specific parts of the input. For the audio tagging task, the proposed attention method can automatically select and attend on the important frames for the targets while ignoring the unrelated parts (e.g., the background noise segments). It can also be regarded as learning a weighting factor on each frame.
The suppression capability against background noise can make the system more robust with the whole chunk as the input.
The attention scheme in this work is conducted based on the convolutional gated recurrent neural network \cite{yong2017ijcnn}.

We also define another {\em \textbf{localization}} module to find the temporal locations when the specific event happens. Localizing the acoustic events occurring in the audio recording would be meaningful given that the labels are in chunk-level rather than frame-level. The training process would be weakly supervised due to the unobserved latent variables, namely the acoustic event locations. This is similar to the process of weakly-supervised image segmentation with only per-image labels \cite{kolesnikov2016seed}. In our previous work \cite{qq2017icassp}, a joint detection-classification model was proposed to detect the locations of acoustic events. However, we have improved the system by introducing an attention model. Furthermore, the feed-forward neural network used in \cite{qq2017icassp}
was inferior to the convolutional gated recurrent neural network (CGRNN) \cite{yong2017ijcnn} which will be introduced in the following sections. In summary, in our framework, {\em \textbf{attention}} is used for global event-independent frame-level feature selection, while the event-dependent {\em \textbf{localization}} is used to find the locations of each event.

The rest work is organized as follows: in Section \ref{sec:cgrnn}, the convolutional gated recurrent neural network (CGRNN) is presented as the basic framework for audio tagging. In section \ref{sec:att_loc}, the proposed attention and localization methods will be illustrated. The experimental setup and results are shown in Section \ref{sec:exp_setup}. Section \ref{sec:conclusion} summarizes the work and foresees the future work.

\section{Chunk-level convolutional gated recurrent neural network (CGRNN)}
\label{sec:cgrnn}
Convolutional gated recurrent neural network was adopted in our previous work \cite{yong2017ijcnn} for audio tagging. However, it only predicted the tags without localizing the acoustic events. Meanwhile it was not trained on the chunk-level but on the 33-frame context window. The chunk-level CGRNN will be briefly presented in this section.

\begin{figure}[t]
	\centering
	\centerline{\includegraphics[width=\columnwidth]{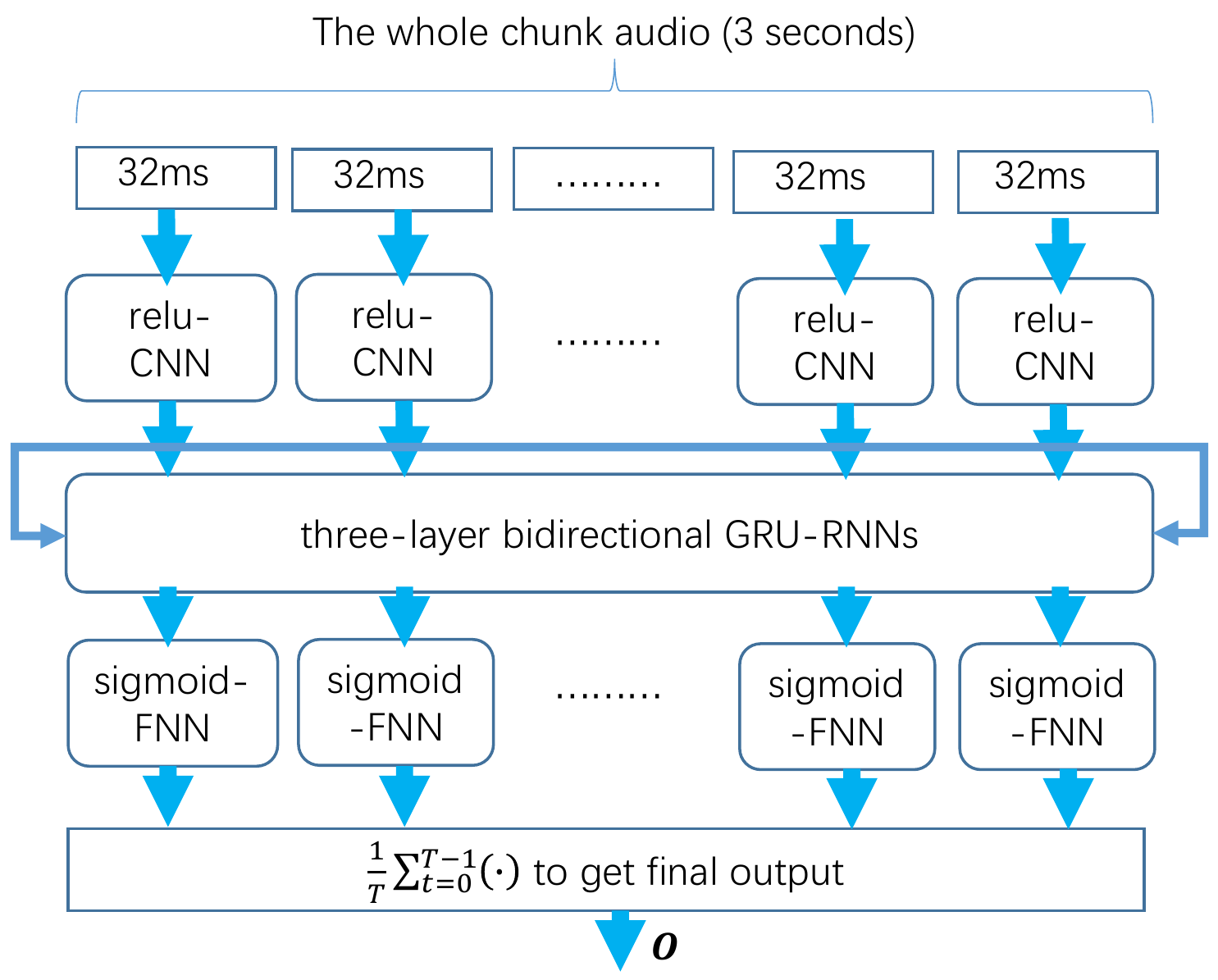}}
	\caption{The framework of the chunk-level convolutional gated recurrent neural network (CGRNN) for audio tagging. Note that the input frames are overlapped by half.}
	\label{fig:cnn_rnn_at}
\end{figure}
The framework of the chunk-level convolutional gated recurrent neural network for audio tagging is shown in Fig. \ref{fig:cnn_rnn_at}. The whole audio chunk is chopped into frames with half overlap. Each frame is fed into a convolutional neural network (CNN) with a large receptive field considering that only one CNN layer is used. The CNN can help to extract more robust features through the max-pooling operations. Rectified Linear Unit (ReLU) is the activation function of CNNs. More details of the CNN configuration could be found in \cite{yong2017ijcnn}. The output activations of each frame from the CNNs are fed into the following gated recurrent unit (GRU) based recurrent neural network (RNN). GRU \cite{cho2014learning} is an alternative structure to the long-short term memory (LSTM) and the GRU was demonstrated to be better than LSTM in some tasks \cite{chung2014empirical}. The bidirectional GRU-RNN can well model the long-term pattern along the whole chunk \cite{yong2017ijcnn}. The details of GRU-RNN can also be found in \cite{yong2017ijcnn}. Then three-layer GRU-RNNs are followed by one-layer feed-forward neural network (FNN) and the activation function is Sigmoid. The audio tagging is a multi-label task which means several acoustic events could happen simultaneously. Hence the output activation function should be sigmoid. Finally each frame can generate one prediction for the audio tags. Their results should be averaged together to obtain the final predictions. The errors by comparing the predictions with the reference tags can be back-propagated (BP) \cite{werbos1990backpropagation} to update the weights. 

Binary cross-entropy is used as the loss function in our work, since it was demonstrated to be better than the mean squared error in \cite{xu2017trans} for labels with zero or one values. The loss can be defined as:
\begin{equation}
E=-\sum_{n=1}^{N}(\textbf{P}_{n}\text{log}{\textbf{O}}_{n}+(1-\textbf{P}_{n})\text{log}(1-{\textbf{O}}_n))
\label{eq:DNNerrors_bce}
\end{equation}
\begin{equation}
{\textbf{O}}=\frac{1}{T}\sum_{t=0}^{T-1}(1+\text{exp}(-\textbf{S}_t))^{-1}
\label{eq:DNN_hidden_sigmoid}
\end{equation}
where $E$ is the binary cross-entropy, ${\textbf{O}}_n$ and $\textbf{P}_{n}$ denote the estimated and reference tag vector at sample index $n$, respectively. The bunch size is represented by $N$. The DNN linear output is defined as $\textbf{S}_t$ at $t$-th frame before the sigmoid activation function is applied. $T$ denotes the total number of frames in the whole audio chunk. Adam \cite{kingma2014adam} is used as the stochastic optimization method. 

\section{Proposed attention and localization (ATT-LOC) methods based on CGRNN}
\label{sec:att_loc}
The attention and localization (ATT-LOC) schemes in the CGRNN framework will be introduced in this section.
\subsection{Attention for audio tagging}
\begin{figure}[t]
	\centering
	\centerline{\includegraphics[width=2.5in]{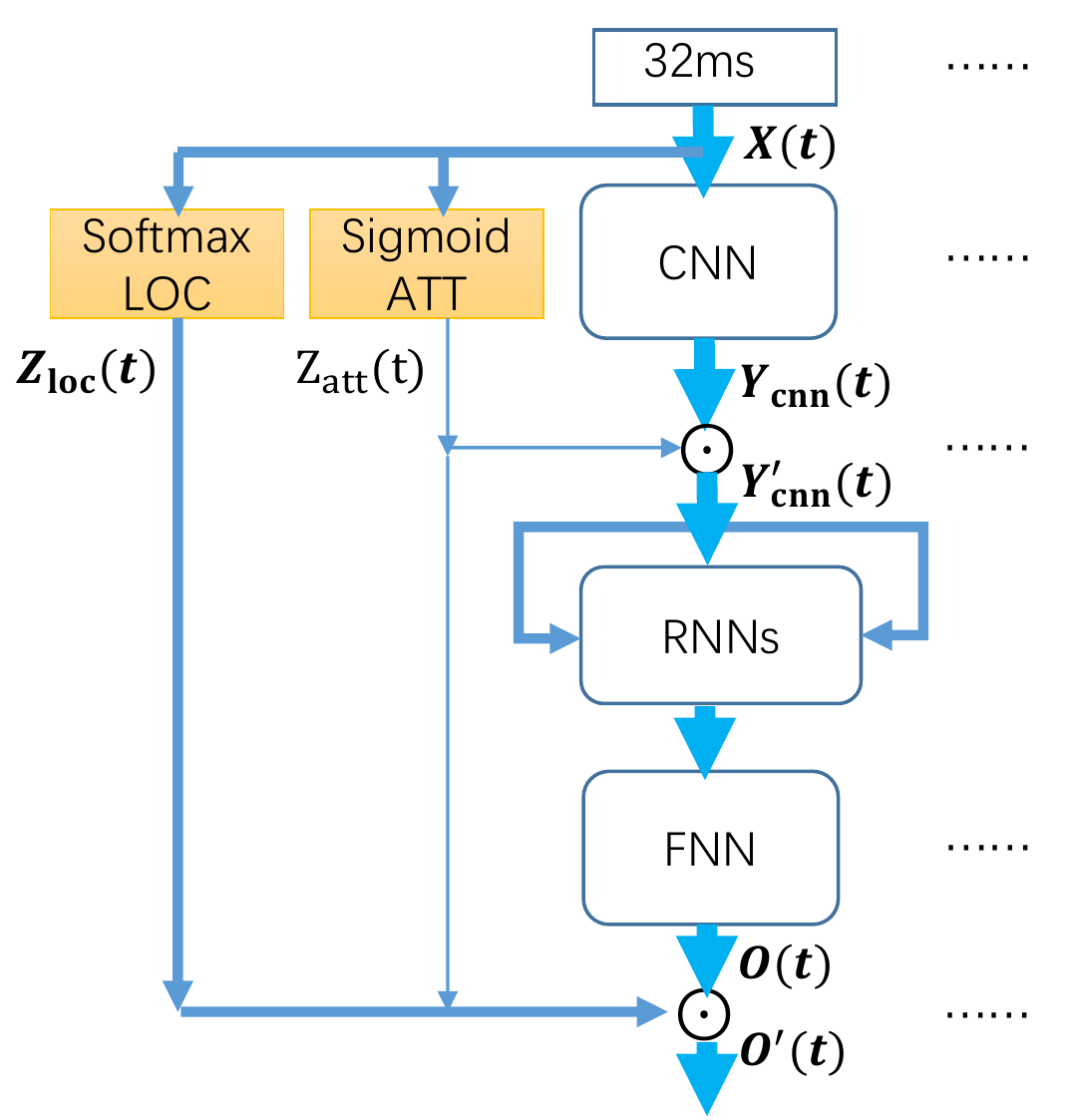}}
	\caption{The diagram of the attention and localization schemes based on CGRNN for audio tagging. ATT denotes the attention module. LOC represents the localization module.}
	\label{fig:att_loc}
\end{figure}
The attention scheme is an additional sigmoid layer with one node output which is shown in Fig. \ref{fig:att_loc}. The predicted attention factor $Z_{\text{att}}(t)$ at the $t$-th frame indicates the importance of the current frame for the final labels. It is learned as,
\begin{equation}
Z_{\text{att}}(t)=\sigma(\textbf{W}_{\text{att}}*\textbf{X}(t)+{b}_{\text{att}})
\label{eq:att}
\end{equation}
where $\textbf{X}(t)$ is the input feature at the $t$-th frame. $\sigma$ is the \textit{Sigmoid} function. $\textbf{W}_{\text{att}}$ and ${b}_{\text{att}}$ denote the weights and bias of the attention module. As $Z_{\text{att}}(t)$ is the latent variable which should be inferred according to the observed chunk-level tags, only one layer without any hidden layer was designed. Then the predicated attention factor is multiplied with the CNN output to suppress the background noise as following,
\begin{equation}
\textbf{Y}^{\prime}_{\text{cnn}}(t)=Z_{\text{att}}(t)\textbf{Y}_{\text{cnn}}(t) 
\label{eq:att_fea}
\end{equation}
where $\textbf{Y}_{\text{cnn}}(t)$ represents the activations from CNN. 
The attention-weighted feature against the background noise is denoted by $\textbf{Y}^{\prime}_{\text{cnn}}(t)$. This attention-weighting process can select the important frames while suppressing the unrelated frames. Finally, the predicated attention factor is also applied to the final acoustic tag outputs at each frame. It is defined as,
\begin{equation}
\textbf{O}^{\prime}(t)=Z_{\text{att}}(t)\textbf{O}(t)
\label{eq:att_tag}
\end{equation}
where $\textbf{O}(t)$ denotes the tag prediction output at the $t$-th frame. The attention factor $Z_{\text{att}}(t)$ can help to decide its contribution degree at the $t$-th frame for the final chunk-level answer. Hence the weighted output is denoted as $\textbf{O}^{\prime}(t)$. The background noise in the audio recordings leads potentially to the over-fitting problem. While the introduced attention method can alleviate the over-fitting problem especially when the input is chunk-level features. Longer input means more noise was fed into the model. That is reason for a context window input (e.g., 32 frames in \cite{Cakir2016}) used in \cite{Cakir2016,Kong2016,xu2016fully,yong2017ijcnn} without any attention-based feature selection schemes.

\subsection{Temporal localization for each acoustic event}
The proposed attention module is to predict the importance of each frame. However the localization module is to localize the occurred acoustic events in the whole audio chunk. For example, there are seven acoustic tags defined in the audio tagging task. It is meaningful to predict the accurate temporal locations (at frame-level) of the occurred acoustic events. Nonetheless, the training will be difficult due to the availability of only the chunk-level labels rather than the frame-level labels. We called the training process as  a weakly supervised process. The localization method is also shown in Fig. \ref{fig:att_loc}. The localization module is one softmax layer without any hidden layer for tractable learning. Similar to the attention factor calculation, the localization vector $\textbf{Z}_{\text{loc}}(t)$ is calculated as,
\begin{equation}
\textbf{Z}_{\text{loc}}(t)=\lambda(\textbf{W}_{\text{loc}}*\textbf{X}(t)+\textbf{b}_{\text{loc}})
\label{eq:loc}
\end{equation}
where $\lambda$ is the \textit{Softmax} function. $\textbf{Z}_{\text{loc}}(t)$ denotes the localization vector at the $t$-th frame. There are seven acoustic events defined in the audio tagging task. Hence the localization vector $\textbf{Z}_{\text{loc}}(t)$ contains the posterior of each acoustic event, and their posterior sum is equal to one. Then the localization vector $\textbf{Z}_{\text{loc}}(t)$ is multiplied with the model classification output at each frame. Then Eq. (\ref{eq:att_tag}) will be updated as,
\begin{equation}
\textbf{O}^{\prime}(t)=Z_{\text{att}}(t)\textbf{O}(t) \odot \textbf{Z}_{\text{loc}}(t)
\label{eq:att_det_tag}
\end{equation}
where the specific dimension of the localization vector $\textbf{Z}_{\text{loc}}(t)$ corresponds to the specific output node (namely the certain acoustic event) of the model. Therefore, the localization vector $\textbf{Z}_{\text{loc}}(t)$ can predict the locations of each acoustic event along the audio chunk. $\odot$ represents the element-wise multiplication. To get the final acoustic event tag predictions, $\textbf{O}^{\prime}(t)$ should be averaged across the audio chunk to get the final output $\textbf{O}^{\prime\prime}$. $\textbf{O}^{\prime\prime}$ is defined as the weighted average of $\textbf{O}^{\prime}(t)$ as following,
\begin{equation}
\textbf{O}^{\prime\prime}=\frac{\sum_{t=0}^{T-1}\textbf{O}^{\prime}(t)}{\sum_{t=0}^{T-1}\textbf{Z}_{\text{loc}}(t)}
\label{eq:att_det_tag_final}
\end{equation}
where the value of the localization vector $\textbf{Z}_{\text{loc}}(t)$ is ranged from zero to one. The sum of  $\textbf{Z}_{\text{loc}}(t)$ at the $t$-th frame is equal to one. Finally the predictions $\textbf{O}^{\prime\prime}$ and the reference acoustic event tags are compared to calculate the back-propagation error.

\subsection{Relationships between the attention and localization modules}
The attention factor $Z_{\text{att}}(t)$ defined in Eq. (\ref{eq:att}) and the localization vector $\textbf{Z}_{\text{loc}}(t)$ defined in Eq. (\ref{eq:loc}) are actually latent variables at frame-level. The proposed model shown in Fig. \ref{fig:att_loc} can infer their prediction values through the chunk-level observations (or labels). The attention module is necessary for the localization module. The activation function of the localization module is Softmax which indicates that there must be at least one event occurring at each frame. However, this assumption is not always reasonable due to the presence of the background noise frames without any meaningful events occurring. As defined in Eq. (\ref{eq:att_det_tag}), the attention factor would mask the values of the localization vectors to zero if there was nothing happening at certain frames. The localization vectors $\textbf{Z}_{\text{loc}}(t)$ are actually local attention factors while the $Z_{\text{att}}(t)$ is a global attention factor. $Z_{\text{att}}(t)$ can select the important features while suppressing the unrelated information, e.g., the background noise frames. It will help to smooth the mismatch or over-fitting problem between the training chunks and the testing chunks. On the other hand, the local attention $\textbf{Z}_{\text{loc}}(t)$ can find the locations, or can focus on the corresponding features for different acoustic events. Hence, the attention factor $Z_{\text{att}}(t)$ is acoustic event independent while the localization vector $\textbf{Z}_{\text{loc}}(t)$ is event dependent.

\section{Experimental setup and results}
\label{sec:exp_setup}
\subsection{Experimental setup}
The experiments are conducted on the DCASE 2016 audio tagging challenge \cite{dcase_t4}. The audio recordings were made in a domestic environment \cite{christensen2010, foster2015}. The audio data are provided as 4-second chunks at 16kHz sampling rate. There are seven acoustic event tags shown in Table \ref{tab:annotations}. The number of recordings is 4387 for the development set and 816 for the evaluation set. Five-fold sets are configured in the development set.
\begin{table}[h] 
	\centering
	\caption{Seven audio events used as the reference labels.}
	\begin{tabular}{|c|c|}
		\hline
		audio event & Event descriptions \\ \hline
		`b' & Broadband noise \\ 
		`c' & Child speech \\
		`f' & Adult female speech \\
		`m' & Adult male speech \\
		`o' & Other identifiable sounds \\
		`p' & Percussive sound events, \\
		& e.g. footsteps, knock, crash \\
		`v' & TV sounds or Video games \\
		\hline
	\end{tabular}
	\label{tab:annotations}
\end{table}

The parameters of the networks are similar to those in our previous work \cite{yong2017ijcnn}. The CNN has 128 filters with the kernel size equal to 30 \cite{yong2017ijcnn, sainath2015learning}. Mel-Filter banks (MFB) with 40 channels are adopted as the input features. The CNN layer is followed by three bidirectional RNN layers with 128 GRU blocks. One feed-forward layer with 500 ReLU units is finally connected to the 7 sigmoid output units. The attention factor is a 1-dimensional scaler at $t$-th frame. The localization vector at $t$-th frame is 7-dimensional which is bounded to the classification output. For performance evaluation, we use equal error rate (EER) \cite{murphy2012} as the main metric which is also suggested by the DCASE 2016 audio tagging challenge. The source codes for this paper can be downloaded from Github\footnote{\url{https://github.com/yongxuUSTC/att_loc_cgrnn}}. More attention and localization demos can also be found on the web\footnote{\url{https://sites.google.com/view/xuyong/demos/attention_model}}.

We compared our methods with the state-of-the-art systems. Lidy-CNN \cite{Lidy2016} and Cakir-CNN \cite{Cakir2016} won the first and the second prize of the DCASE2016 audio tagging challenge \cite{dcase_t4}. They both used CNN as the classifier. We also compare this method to our previous method CGRNN \cite{yong2017ijcnn} which demonstrated the best performance using the convolutional gated recurrent neural network. Our another previous method, de-noising auto-encoder (DAE) \cite{xu2017trans} based audio tagging, was also used as a baseline.

\subsection{Results and analysis}
\label{ssec:results}
In this sub-section, the localization and attention demos will be shown firstly, then the overall evaluations on the development set and the evaluation set of Task 4 of the DCASE 2016 challenge will be given.

\subsubsection{Predicted localization and attention results}
\begin{figure}[t]
	\centering
	\centerline{\includegraphics[width=\columnwidth]{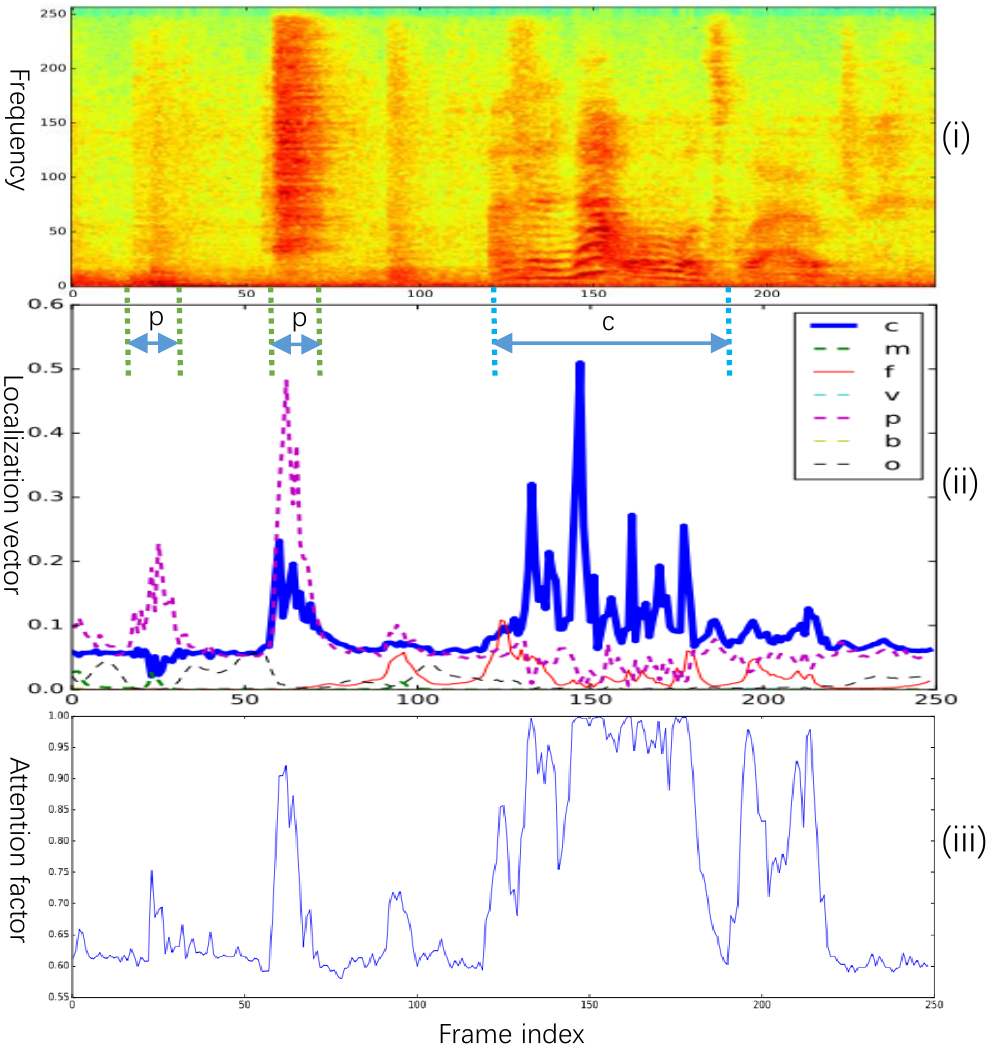}}
	\caption{The logarithmic spectrogram denoted as (i), the predicted localization results denoted as (ii) and the attention factor denoted as (iii) for an audio chunk labeled as ``child speech (c)" and ``percussive sound (p)". The X-axis of the three figures are all in the same frame index. The corresponding audio file can be also auditioned at the demo website.}
	\label{fig:cnn_rnn_at_demos}
\end{figure}

Fig.\ref{fig:cnn_rnn_at_demos} presents the logarithmic spectrogram denoted as (i), the predicted localization results denoted as (ii) and the attention factor denoted as (iii) for an audio chunk ``CR\_lounge\_220110\_0731.s0\_chunk70'' which is labeled as ``child speech (c)" and ``percussive sound (p)". In fact, this audio tagging task only gives the chunk-level labels rather than the frame-level labels. However, the rough locations of the occurred events can be labeled manually to compare with the predictions of the proposed methods. As shown in Fig. \ref{fig:cnn_rnn_at_demos}, two ``percussive sound (p)" sounds (represented by the dashed purple line in the middle figure) are accurately localized. The ``child speech (c)'' segments are also successfully localized. Meanwhile, the predicted posteriors of other events which are not occurring in this chunk are almost suppressed along the whole chunk. The predicted attention factor is shown in the (iii) figure. It can be found that the attention can capture the important segments where related events happen while suppress the contribution of other non-related segments. All of the values of the attention factor are bigger than 0.5. It indicates that the attention scheme tends to keep some information with the adopted Sigmoid activation function in the attention module.
\begin{table}[h] \scriptsize 
	\centering
	\caption{EER results of the proposed method ATT-LOC and the CGRNN \cite{yong2017ijcnn} method on the \textbf{development set} of the Task 4 of DCASE 2016 challenge, across the seven audio event tags.}
	\begin{tabular}{l|cccccccc}
		\hline
		Dev-set & c     & m     & f     & v     & p     & b     & o     & ave \\
		\hline
		{ CGRNN \cite{yong2017ijcnn}} & 0.14  & \textbf{0.09} & 0.17  & \textbf{0.02} & 0.13  & 0.04  & 0.24  & 0.12 \\
		{ ATT-LOC} & \textbf{0.10} & 0.10  & \textbf{0.16} & 0.03  & \textbf{0.11} & \textbf{0.03} & \textbf{0.22} & \textbf{0.11} \\
		\hline
	\end{tabular}%
	
	\label{tab:dev_data}
\end{table}

The model is weakly-supervised with only chunk-level labels. Why does it still have the ability to predict the detailed locations of the occurring audio events? There are seven Softmax output nodes in the localization module (shown in Fig. \ref{fig:att_loc}), and each node of the localization module is specifically connected to one of the seven classification output nodes. Each output node is corresponding to a specific audio event or tag. Therefore, the latent locations of the occurring audio events can be inferred through the chunk-level training.

\subsubsection{Overall evaluations}
\begin{table}[h] \scriptsize 
	\centering
	\caption{EER comparisons on the \textbf{evaluation set} among several newly proposed methods across the seven audio event tags.}
\begin{tabular}{l|cccccccc} 
	\hline
	Eval-set & c     & m     & f     & v     & p     & b     & o     & ave \\
	\hline
	{ Cakir-CNN \cite{Cakir2016}} & 0.25  & 0.16  & 0.25  & 0.03  & 0.21  & 0.02  & 0.26  & 0.17 \\
	{ Lidy-CNN \cite{Lidy2016}}  & 0.21  & 0.18  & 0.21  & 0.04  & 0.17  & 0.03  & 0.32  & 0.17 \\
	{ DAE-DNN \cite{xu2017trans}} & 0.21  & 0.15  & 0.21  & \textbf{0.02} & 0.18  & 0.01  & 0.26  & 0.15 \\
	{ CGRNN \cite{yong2017ijcnn}} & 0.17  & 0.16  & 0.18  & 0.03  & 0.15  & \textbf{0.00} & 0.24  & 0.13 \\
	{ ATT-LOC} & \textbf{0.09} & \textbf{0.14} & \textbf{0.17} & 0.03  & \textbf{0.12} & 0.01  & \textbf{0.24} & \textbf{0.11} \\
	\hline
\end{tabular}%
	\label{tab:eval_data}
\end{table}

In Table \ref{tab:dev_data}, we firstly verify the effectiveness of the proposed method by comparing it with the most competitive system proposed recently in \cite{yong2017ijcnn}. The attention and localization method can get slightly smaller EER. Then in Table \ref{tab:eval_data}, full comparisons are conducted among several newly proposed methods on the evaluation set of the audio tagging task. Compared with the CGRNN method \cite{yong2017ijcnn}, the proposed attention and localization method reduces the EER from 0.13 to 0.11 on average. It is found that the proposed method can get significantly improvement for the ``child speech (c)'' audio event both on the development set and the evaluation set. The ``child speech (c)'' audio event is the most frequent event occurring in the whole dataset. The attention and localization scheme performs better in detecting the long-term pattern of the ``child speech''.

\section{Conclusions}
\label{sec:conclusion}
In this paper, we proposed a new audio tagging method based on our previous work using CGRNN \cite{yong2017ijcnn} by introducing the attention and localization scheme. It not only can reduce the overall EER on the evaluation set from 0.13 to 0.11, but also can infer the latent temporal locations of each occurring event in a weakly-supervised mode. This weakly-supervised method to predict the locations of events with only the chunk-level label is useful in the real-world application scenario. It is much easier to get the chunk-level labels considering that the frame-level labels are time-consuming and less accurate under the human manual effort. Hence, in the near future, we will evaluate our proposed method on large data sets, such as the Yahoo Flickr Creative Commons 100 Million (YFCC100m) dataset \cite{thomee2016yfcc100m}, YouTube-8M dataset \cite{abu2016youtube} and Google audio set \cite{audioset}.

\section{Acknowledgements}
This work was supported by the Engineering and Physical Sciences Research Council (EPSRC) of the UK under the grant EP/N014111/1. Qiuqiang Kong is partially supported by China scholarship council (CSC).

\newpage
\bibliographystyle{IEEEtran}

\bibliography{mybib}

\begin{thebibliography}{10}
\providecommand{\url}[1]{#1}
\csname url@samestyle\endcsname
\providecommand{\newblock}{\relax}
\providecommand{\bibinfo}[2]{#2}
\providecommand{\BIBentrySTDinterwordspacing}{\spaceskip=0pt\relax}
\providecommand{\BIBentryALTinterwordstretchfactor}{4}
\providecommand{\BIBentryALTinterwordspacing}{\spaceskip=\fontdimen2\font plus
\BIBentryALTinterwordstretchfactor\fontdimen3\font minus
  \fontdimen4\font\relax}
\providecommand{\BIBforeignlanguage}[2]{{%
\expandafter\ifx\csname l@#1\endcsname\relax
\typeout{** WARNING: IEEEtran.bst: No hyphenation pattern has been}%
\typeout{** loaded for the language `#1'. Using the pattern for}%
\typeout{** the default language instead.}%
\else
\language=\csname l@#1\endcsname
\fi
#2}}
\providecommand{\BIBdecl}{\relax}
\BIBdecl

\bibitem{giannoulis2013detection}
D.~Giannoulis, E.~Benetos, D.~Stowell, M.~Rossignol, M.~Lagrange, and M.~D.
  Plumbley, ``Detection and classification of acoustic scenes and events: an
  {IEEE AASP} challenge,'' in \emph{2013 IEEE Workshop on Applications of
  Signal Processing to Audio and Acoustics}, pp. 1--4.

\bibitem{stowell2015detection}
D.~Stowell, D.~Giannoulis, E.~Benetos, M.~Lagrange, and M.~D. Plumbley,
  ``Detection and classification of acoustic scenes and events,'' \emph{IEEE
  Transactions on Multimedia}, vol.~17, no.~10, pp. 1733--1746, 2015.

\bibitem{dcase2016}
\url{http://www.cs.tut.fi/sgn/arg/dcase2016/}.

\bibitem{dcase_t4}
\url{http://www.cs.tut.fi/sgn/arg/dcase2016/task-audio-tagging}.

\bibitem{foster2015}
P.~Foster, S.~Sigtia, S.~Krstulovic, J.~Barker, and M.~D. Plumbley,
  ``{CHiME}-home: A dataset for sound source recognition in a domestic
  environment,'' in \emph{IEEE Workshop on Applications of Signal Processing to
  Audio and Acoustics}, 2015, pp. 1--5.

\bibitem{kumar2016audio}
A.~Kumar and B.~Raj, ``Audio event detection using weakly labeled data,'' in
  \emph{Proceedings of the 2016 ACM on Multimedia Conference}.\hskip 1em plus
  0.5em minus 0.4em\relax ACM, 2016, pp. 1038--1047.

\bibitem{Foster2016}
\BIBentryALTinterwordspacing
P.~Foster and T.~Heittola, ``{DCASE}2016 baseline system,'' \emph{IEEE AASP
  Challenge on Detection and Classification of Acoustic Scenes and Events
  (DCASE 2016) challenge}. [Online]. Available:
  \url{https://github.com/pafoster/dcase2016_task4/tree/master/baseline}
\BIBentrySTDinterwordspacing

\bibitem{Yun2016}
\BIBentryALTinterwordspacing
S.~Yun, S.~Kim, S.~Moon, J.~Cho, and T.~Kim, ``Discriminative training of {GMM}
  parameters for audio scene classification and audio tagging,''
  \textit{DCASE2016} \textit{Challenge}, Tech. Rep., 2016. [Online]. Available:
  \url{http://www.cs.tut.fi/sgn/arg/dcase2016/documents/challenge_technical_reports/Task4/Yun_2016_task4.pdf}
\BIBentrySTDinterwordspacing

\bibitem{xu2016fully}
Y.~Xu, Q.~Huang, W.~Wang, P.~Jackson, and M.~Plumbley, ``Fully dnn-based
  multi-label regression for audio tagging,'' in \emph{Proceedings of the
  Detection and Classification of Acoustic Scenes and Events 2016 Workshop
  (DCASE2016)}, 2016, pp. 110--114.

\bibitem{Kong2016}
\BIBentryALTinterwordspacing
Q.~Kong, I.~Sobieraj, W.~Wang, and M.~D. Plumbley, ``Deep neural network
  baseline for {DCASE} challenge 2016,'' \textit{DCASE2016 Challenge}, Tech.
  Rep., 2016. [Online]. Available:
  \url{http://www.cs.tut.fi/sgn/arg/dcase2016/documents/challenge_technical_reports/Task4/Kong_2016_task4.pdf}
\BIBentrySTDinterwordspacing

\bibitem{Cakir2016}
\BIBentryALTinterwordspacing
E.~Cakir, T.~Heittola, and T.~Virtanen, ``Domestic audio tagging with
  convolutional neural networks,'' \textit{DCASE2016 Challenge}, Tech. Rep.,
  2016. [Online]. Available:
  \url{http://www.cs.tut.fi/sgn/arg/dcase2016/documents/challenge_technical_reports/Task4/Cakir_2016_task4.pdf}
\BIBentrySTDinterwordspacing

\bibitem{Lidy2016}
\BIBentryALTinterwordspacing
T.~Lidy and A.~Schindler, ``{CQT}-based convolutional neural networks for audio
  scene classification and domestic audio tagging,'' \textit{DCASE2016
  Challenge}, Tech. Rep., 2016. [Online]. Available:
  \url{http://www.cs.tut.fi/sgn/arg/dcase2016/documents/challenge_technical_reports/Task4/Lidy_2016_task4.pdf}
\BIBentrySTDinterwordspacing

\bibitem{yong2017ijcnn}
Y.~Xu, Q.~Kong, Q.~Huang, W.~Wang, and M.~D. Plumbley, ``Convolutional gated
  recurrent neural network incorporating spatial features for audio tagging,''
  in \emph{2017 IEEE International Joint Conference on Neural Networks (IJCNN
  2017)}.

\bibitem{bahdanau2016end}
D.~Bahdanau, J.~Chorowski, D.~Serdyuk, P.~Brakel, and Y.~Bengio, ``End-to-end
  attention-based large vocabulary speech recognition,'' in \emph{Proceedings
  of ICASSP}, 2016, pp. 4945--4949.

\bibitem{chorowski2015attention}
J.~K. Chorowski, D.~Bahdanau, D.~Serdyuk, K.~Cho, and Y.~Bengio,
  ``Attention-based models for speech recognition,'' in \emph{{A}dvances in
  {N}eural {I}nformation {P}rocessing {S}ystems}, 2015, pp. 577--585.

\bibitem{mnih2014recurrent}
V.~Mnih, N.~Heess, A.~Graves \emph{et~al.}, ``Recurrent models of visual
  attention,'' in \emph{{A}dvances in {N}eural {I}nformation {P}rocessing
  {S}ystems}, 2014, pp. 2204--2212.

\bibitem{bahdanau2014neural}
D.~Bahdanau, K.~Cho, and Y.~Bengio, ``Neural machine translation by jointly
  learning to align and translate,'' \emph{arXiv preprint arXiv:1409.0473},
  2014.

\bibitem{xu2015show}
K.~Xu, J.~Ba, R.~Kiros, K.~Cho, A.~C. Courville, R.~Salakhutdinov, R.~S. Zemel,
  and Y.~Bengio, ``Show, attend and tell: Neural image caption generation with
  visual attention.'' in \emph{{ICML}}, vol.~14, 2015, pp. 77--81.

\bibitem{kolesnikov2016seed}
A.~Kolesnikov and C.~H. Lampert, ``Seed, expand and constrain: Three principles
  for weakly-supervised image segmentation,'' in \emph{European Conference on
  Computer Vision}, 2016, pp. 695--711.

\bibitem{qq2017icassp}
Q.~Kong, Y.~Xu, W.~Wang, and M.~D. Plumbley, ``A joint detection-classification
  model for audio tagging of weakly labelled data,'' \emph{Proceedings of
  ICASSP}, 2017.

\bibitem{cho2014learning}
K.~Cho, B.~Van~Merri{\"e}nboer, C.~Gulcehre, D.~Bahdanau, F.~Bougares,
  H.~Schwenk, and Y.~Bengio, ``Learning phrase representations using {RNN}
  encoder-decoder for statistical machine translation,'' \emph{Conference on
  Empirical Methods in Natural Language Processing (EMNLP 2014)}, 2014.

\bibitem{chung2014empirical}
J.~Chung, C.~Gulcehre, K.~Cho, and Y.~Bengio, ``Empirical evaluation of gated
  recurrent neural networks on sequence modeling,'' \emph{arXiv preprint
  arXiv:1412.3555}, 2014.

\bibitem{werbos1990backpropagation}
P.~J. Werbos, ``Backpropagation through time: What it does and how to do it,''
  \emph{Proceedings of the IEEE}, vol.~78, no.~10, pp. 1550--1560, 1990.

\bibitem{kingma2014adam}
D.~Kingma and J.~Ba, ``Adam: {A} method for stochastic optimization,''
  \emph{arXiv preprint arXiv:1412.6980}, 2014.

\bibitem{christensen2010}
H.~Christensen, J.~Barker, N.~Ma, and P.~Green, ``The {CHiME} corpus: a
  resource and a challenge for computational hearing in multisource
  environments,'' in \emph{Proceedings of Interspeech}, 2010, pp. 1918--1921.

\bibitem{sainath2015learning}
T.~N. Sainath, R.~J. Weiss, A.~Senior, K.~W. Wilson, and O.~Vinyals, ``Learning
  the speech front-end with raw waveform {CLDNN}s,'' in \emph{Proceedings of
  Interspeech}, 2015.

\bibitem{murphy2012}
K.~P. Murphy, \emph{Machine Learning: A Probabilistic Perspective}.\hskip 1em
  plus 0.5em minus 0.4em\relax MIT Press, 2012.

\bibitem{xu2017trans}
Y.~Xu, Q.~Huang, W.~Wang, , P.~Foster, S.~Sigtia, P.~Jackson, and M.~Plumbley,
  ``Unsupervised feature learning based on deep models for environmental audio
  tagging,'' in \emph{{IEEE}/{ACM} Trans. on audio, speech and language
  processing}, 2017.

\bibitem{thomee2016yfcc100m}
B.~Thomee, D.~A. Shamma, G.~Friedland, B.~Elizalde, K.~Ni, D.~Poland, D.~Borth,
  and L.-J. Li, ``{YFCC100M}: The new data in multimedia research,''
  \emph{Communications of the ACM}, vol.~59, no.~2, pp. 64--73, 2016.

\bibitem{abu2016youtube}
S.~Abu-El-Haija, N.~Kothari, J.~Lee, P.~Natsev, G.~Toderici, B.~Varadarajan,
  and S.~Vijayanarasimhan, ``You{T}ube-8{M}: A large-scale video classification
  benchmark,'' \emph{arXiv preprint arXiv:1609.08675}, 2016.

\bibitem{audioset}
J.~F. Gemmeke, D.~P.~W. Ellis, D.~Freedman, A.~Jansen, W.~Lawrence, R.~C.
  Moore, M.~Plakal, and M.~Ritter, ``Audio set: An ontology and human-labeled
  dataset for audio events,'' in \emph{Proceedings of ICASSP}, 2017.

\end{thebibliography}


\end{document}